\documentclass[twocolumn,showpacs,preprintnumbers,amsmath,amssymb]{revtex4}

\usepackage{graphicx}
\usepackage{dcolumn}
\usepackage{bm}
\usepackage{longtable}

\def\gsim {\mbox{\hbox{ \lower-.6ex\hbox{$>$}
\kern-1.12em \lower.5ex\hbox{$\sim$}\kern+.35em}}}
\def\lsim {\mbox{\hbox{ \lower-.6ex\hbox{$<$}
\kern-1.12em \lower.5ex\hbox{$\sim$}\kern+.35em}}}

\newcommand{\eqn}{eq. }
\newcommand{\eqns}{eqs. }
\newcommand{\fig}{Figure }
\newcommand{\rrf}{reference }
\newcommand{\sect}{Section }

\begin{document}


\title{Equilibrium Charge Distribution on Annealed Polyelectrolytes}

\author{Tiziano Zito}
\author{Christian Seidel}%
 \thanks{to whom correspondence should be addressed}
 \email[E-mail address: ]{seidel@mpikg-golm.mpg.de}
\affiliation{%
Max-Planck-Institut f\"ur Kolloid- und Grenzfl\"achenforschung\\
Am M\"uhlenberg, D-14476 Golm, Germany
}%
\altaffiliation{mailing address: D-14424 Potsdam, Germany}


\date{\today}

\begin{abstract}
Monte Carlo simulations are used to study the non-uniform equilibrium
charge distribution along a single annealed polyelectrolyte chain
under $\theta$ solvent conditions and with added salt. Within a range of
the order of the Debye length charge accumulates at chain ends
while a slight charge depletion appears in the central part of the
chain. The simulation results are compared with theoretical
predictions recently given by Castelnovo {\it et al}. In the parameter range
where the theory can be applied we find almost perfect quantitative
agreement.
\end{abstract}

\pacs{61.25.Hq, 36.20.-r, 5.20.-y}


\maketitle

\section{Introduction}

The term polyelectrolyte is employed for the wide field of macromolecules
which contain dissociable subunits (see, e.g., \cite{seid94b,barr95} and
references therein). With respect to different dissociation behavior
one can distinguish between strong and weak polyelectrolytes
\cite{seid94b} or between quenched and annealed ones
\cite{raph90}. So-called strong polyelectrolytes, polysalts as
e.g. Na-polystyrene  sulfonate, dissociate completely in the total pH
range accessible by experiment. The total charge as well as its
specific distribution along the chain is solely imposed by chemistry,
i.e., by polymer synthesis. That is why such polyelectrolytes are also
called quenched. On the other hand, weak polyelectrolytes represented
by polyacids and polybases dissociate only in a rather limited pH
range. The total charge of the chain is not fixed but it can be tuned
by changing the pH of the solution. Because of dissociation and
recombination of ion pairs along the chain one expects spatial and/or
temporal fluctuations in the local degree of dissociation. Such
titrating polyelectrolytes exhibit an annealed inhomogeneous charge
distribution. A pronounced charge accumulation appears at chain ends
because there are fewer neighbors for the charges to interact with and
the penalty in energy is therefore reduced. Although, at the level of
scaling laws describing the statistical properties of polymer chains,
the local charge distribution has only a weak effect on  numerical
pre-factors \cite{barr95} the extra degree of freedom for the charges
leads to new and non trivial features.  The charge inhomogeneity  can
have a strong impact on processes dominated by end-effects, such as
the self-assembly of weakly charged linear micelles \cite{vdS97} and
adsorption on charged surfaces \cite{fleer93}. For end-grafted weak
polyelectrolytes, a rather unusual regime has been obtained where the
chain stretching (brush thickness) depends non-monotonously on salt
concentration \cite{israels94} and grafting density
\cite{zhul95}. This is mainly due to the fact that 
the net charge of a chain as well as its distribution along the chain
is not fixed but depends on its local environment.

The dissociation of a low molecular acid (HA) in an aqueous medium
is given by the equilibrium reaction
\[ \text{HA}\,+\,\text{H}_2\text{O}\, \rightleftharpoons
\,\text{H}_3\text{O}^+ \,+\,\text{A}^- ,\]
or more simply by
\[ \text{HA}\, \rightleftharpoons
\,\text{H}^+ \,+\,\text{A}^-. \]
The law of mass action yields the equilibrium constant
\begin{equation}
  \label{eq:KA}
  K_a = \frac{\left[ \text{H}^+\right]\,\left[
  \text{A}^-\right]}{\left[ \text{HA}\right]},
\end{equation}
where $\left[\text{A}^-\right]$, $\left[ \text{HA}\right]$, $\left[
  \text{H}^+\right]$ are the (monomolar) concentrations of dissociated and
  undissociated acid and dissociated hydrogen, respectively. 
Using the standard notation $\text{pH} = - \text{log}_{10}\left[
  \text{H}^+\right]$, \mbox{$\text{p}K_a = - \text{log}_{10}\left[
  K_a\right]$} and defining the degree of dissociation by 
\begin{equation}
  \label{eq:degree}
  f = \frac{\left[\text{A}^-\right]}{\left[
  \text{HA}\right]\,+\,\left[\text{A}^-\right]}, 
\end{equation}
\eqn(\ref{eq:KA}) gives the well-known relation between
the pH of the solution and degree of dissociation $f$ of a simple acid
\begin{equation}
  \label{eq:pKa}
  \text{pH} = \text{p}K_a \,+\,\text{log}_{10}\left(\frac{f}{1-f}\right).
\end{equation}
The dissociation behavior of polyacids can be described in a similar
way, but the resulting $\text{p}K_a$ value is now an apparent one
(in the physico-chemical literature denoted by
$\text{p}K_{app}$) \cite{katch47}. In contrast to low-molecular-weight
acids, the charged groups of polyacids are linked together along the
chain. Therefore, the dissociation of one acid group is correlated in a
complex way to the position and the number of other charged groups of
the chain resulting in a masking of the intrinsic $\text{p}K_a^0$  of a
(polyelectrolyte) monomer. The corresponding relation can be written
\cite{raph90,over48}  
\begin{equation}
  \label{eq:pKapoly}
  \text{pH} = \text{p}K_a^0
  \,+\,\text{log}_{10}\left(\frac{f}{1-f}\right)\,+\,\frac{1}{N\,k_B
  T}\,\frac{\partial F_{el}}{\partial f},
\end{equation}
with  $F_{el}$ being the electrostatic free energy of the
polyelectrolyte chain and $N$ is the chain length. Introducing the
chemical potential  
\begin{equation}
  \label{eq:mu}
  \mu (f) =  k_B T
  \,\text{log}_{10}\left(\frac{f}{1-f}\right)\,+\,\frac{1}{N}\,\frac{\partial  
  F_{el}}{\partial f},
\end{equation}
\eqn (\ref{eq:pKapoly}) can be written
\begin{equation}
  \label{eq:pHmu}
  \text{pH} = \text{p}K_a^0 \,+\,\frac{1}{k_B T}\,\mu(f).
\end{equation}
Clearly, the chemical potential has two contributions: (i) an
entropic one related to the mixing of charged and non-charged
groups along the chain and (ii) an electrostatic one related
to the interaction with charged groups forming the local charge
environment of an ionizable site. For good and $\theta$
solvents, the electrostatic contribution $\mu_{el}(f)=N^{-1}\,\partial  
  F_{el}/\partial f$ is an increasing monotonic function of $f$. For
poor solvents, Raphael and Joanny \cite{raph90} found a non-monotonic
variation of $\mu_{el}(f)$, and thus of $\mu(f)$, with $f$ which results in
a first order phase transition from a collapsed weakly charged
conformation to an extended strongly charged state.

The particular behavior of weak polyelectrolytes has attracted
considerable interest in experimental \cite{mandel88,koetz90,Tirta97},
theoretical \cite{raph90,Joe96,berg97,boru00,castel00} 
and simulation \cite{berg97,reed92,sassi92,Ull96a,Ull96b}
studies. Because a usual experimental approach to characterizing weak
polyelectrolytes is to perform titration experiments much effort has
been done to understand the titration curves, i.e., the dependence of
the degree of ionization (or the degree of neutralization in the
titration experiment) on the pH of the solution. The inhomogeneous
charge distribution has been first studied by numerical simulation
\cite{berg97,Ull96a}. For not too large mean charge densities $\left<
  f \right> $ and screening lengths $\lambda_D$, a quite good
agreement between simulation data and the predictions of a linearized density
functional approach has been obtained in the case of rod-like
polyelectrolytes \cite{vdS97,berg97}. Recently a generalization of the theory
to the case of flexible chains has been given by Castelnovo {\it et
  al.} \cite{castel00}. Qualitatively they found the result obtained
in our previous simulations: a charge accumulation at chain
ends. However, a quantitative comparison between theory and simulation
data was not possible because the chains considered by Berghold {\it at al.}
\cite{berg97} are not in the asymptotic regime where the approach used
in theory can be applied. A similar end-effect has been demonstrated
in simulations of quenched strongly charged polyelectrolytes with explicit
counterions: around the ends the counterion distribution is
significantly different from that around the inner part of the chain
\cite{limb01}. However, although the resulting effective charge looks
quite similar to that of an annealed polyelectrolyte a quantitative
comparison with the corresponding theoretical predictions is not
possible due to some important differences between the 
two systems. So the theoretical predictions given by Castelnovo {\it et
  al.} are still waiting for a quantitative verification.

In the present paper we study the non-uniform charge distribution on
annealed weakly charged polyelectrolytes in a $\theta$ solvent by
(semi-)grand canonical Monte Carlo simulation. To be close to the
theoretical model we consider chains where neighboring monomers are
bound by harmonic springs. In addition, charged monomers interact with
a Debye-H\"uckel potential, the screening length of which is tuned over
the range where the theory is valid. In this case we observe not only
qualitative agreement between simulation data and theory, but indeed a
quite good quantitative one. Although the theoretical single chain
problem and its solution in a restricted parameter range may seem
rather academic it is a first step towards a more precise
understanding of unusual collective properties of, e.g., annealed
polyelectrolyte brushes and stars. 
 
The outline of the paper is as follows. In \sect \ref{sec:theory}  we
present the main theoretical predictions. The model and the method we
use in the simulation are described in \sect \ref{sec:simmodel}. In
\sect \ref{sec:results} we discuss the results and compare theory and
simulation. Finally, a brief summary and our conclusions can be found
in \sect \ref{sec:conclu}.
\section{Theory}
\label{sec:theory}
The free energy of an annealed, fully-stretched polyelectrolyte chain
($N$ monomers of size $b$) in a salty solution can be written
\cite{Hansen86} 
\begin{eqnarray}
      \label{eq:F_annrod}
  \frac{F}{k_B T} &=& \int_{-N/2}^{N/2}  ds  \,\Bigg\{  f(s) \left[ \log
  \big(f(s)\big)\, -\, 1\right] -\, \mu \, f(s) \,\nonumber \\
  & & \hspace*{-1.2cm}+ \,
  \frac{\lambda_B}{2}  \int_{-N/2}^{N/2} ds' \,f(s) f(s')\, \frac{\exp
  (-|s-s'|\,b/\lambda_D )}{|s-s'|\,b} \Bigg\}, 
\end{eqnarray}
where $f(s)$ is the local charge distribution along the chain. The
Bjerrum length, which sets the strength of electrostatic interactions,
is given by \mbox{$\lambda_B = e^2 / 4\pi \varepsilon_0 \varepsilon
  k_B T$} with $e$ being the elementary charge and $\varepsilon$ is the
dielectric constant of the medium. Assuming a mono-valent low-molecular
electrolyte of concentration $c_s$, Debye screening length $\lambda_D$
and Bjerrum length are related by $\lambda_D = (8\pi\lambda_B c_s
)^{-1/2}$. The first term of \eqn (\ref{eq:F_annrod}) is the
entropy of an unidimensional ideal gas, the second term fixes the
charge on the chain by the chemical potential $\mu$ and the
third term represents the electrostatic interaction between the
charges on the chain. Note that one can justify the use of a Debye-H\"uckel
potential only if the average charge density $\left<f\right>/\,b$
fulfills certain conditions: (i) $\left<f\right>$ has to be small to
make non-linear effects unimportant, i.e.,  
\begin{equation}
  \label{eq:manning_rod}
  \left<f\right> \frac{\lambda_B}{b}\, =\, \left<f\right> u\, <\, 1,
\end{equation}
where we introduced the dimensionless parameter
$u=\lambda_B/b$, and  (ii) $\left<f\right>$ has to be large to
make sure that a sufficiently large number of charges interact
simultaneously, i.e.,   
\begin{equation}
  \label{eq:Debye_restrict}
  \left<f\right> \frac{\lambda_D}{b}\, >\, 1.
\end{equation}
Note that $\left<f\right> \lambda_B/b$ is the so-called Manning
parameter for the condensation of counterions on a partially charged
rigid rod.

Minimizing \eqn (\ref{eq:F_annrod}) the equilibrium charge density
distribution on an annealed, fully stretched polyelectrolyte chain, up to
first order in $\left< f \right> \lambda_B/b $, was found to be  \cite{berg97}
\begin{eqnarray}
  \label{eq:chargedist_rod}
  \frac{f (s)}{\left< f \right>} 
   &=&  1  + 
   \left< f \right> \frac{\lambda_B}{b} 
  \\   & &\hspace*{-1cm}\times \!\left\{
   E_1\!\left[\!\left(\!\frac{N}{2}+s\!\right)\!\frac{b}{\lambda_D} \!\right]
    +  
   E_1\!\left[\!\left(\!\frac{N}{2}-s\!\right)\!\frac{b}{\lambda_D} \!\right] 
   -  2 \frac{\lambda_D}{N b}
   \right\},
\nonumber
\end{eqnarray}
with the exponential integral $E_1(x) = \int_x^\infty dt\,t^{-1}\exp
(-t)$ \cite{Abram65}. Equation (\ref{eq:chargedist_rod}) gives a
charge accumulation at the ends of the rod within a region of the
order of the screening length. For a wide parameter range, the
theoretical prediction was quantitatively confirmed by Monte Carlo
simulations  \cite{berg97}. 

In order to study the charge distribution on a weakly charged, flexible
polyelectrolyte chain Castelnovo {\it et al.} \cite{castel00}
generalized \eqn (\ref{eq:F_annrod}) by including the
entropy of a freely jointed chain. In the $\theta$ solvent
case the free energy can be written 
\begin{eqnarray}
  \label{eq:F_annflex}
  \frac{F[{\bf r}_0(s)]}{k_B T} &=& \!\int_{-N/2}^{N/2}  ds  \,\Bigg\{
  f(s) \left[ \log 
  \big(f(s)\big)\, -\, 1\right] -\, \mu \, f(s) \nonumber \\
  &+& 
  \frac{3}{2b^2} \left(\frac{d{\bf r}_0(s)}{ds} \right)^2
   + \,\frac{\lambda_B}{2}  \int_{-N/2}^{N/2} ds' \,f(s) f(s')\,\,\nonumber\\
  & & \hspace*{.9cm}\times \,\frac{\exp
  \left[-|{\bf r}_0(s)-{\bf r}_0(s')|/\lambda_D \right]}{|{\bf r}_0(s)-{\bf
  r}_0(s')|} \Bigg\},   
\end{eqnarray}
where ${\bf r}_0(s)$ denotes the so-called classical path of the
polymer (the most probable conformation) which follows from the ``chain
under tension model'' \cite{barrat93}. An important length scale which comes
now into play is the electrostatic blob size $\xi$. From scaling
arguments, i.e., by equating the (unscreened) electrostatic
interaction of the (quenched) charges inside one blob with $k_B T$, one gets
\begin{equation}
  \label{eq:xi_scaling}
  \xi_{\mbox{\scriptsize scaling}} = \frac{b}{\left( u f^2
  \right)^{\scriptstyle 1/3}}\,. 
\end{equation}
One blob contains $g = (\xi_{\mbox{\scriptsize scaling}}/b)^{2}$ monomers and a
chain of $N$ monomers consists of 
\begin{equation}
  \label{eq:blobnumber}
  n_b = N/g = 
    N\,( u f^2)^{2/3} 
\end{equation}
electrostatic blobs. In order to reach the asymptotic regime one has
to ensure that \mbox{$n_b \gg 1$.} For a chain of finite length, however, the
situation becomes even more complex. The electrostatic blob size
varies along the chain resulting in a trumpet-like conformation with
the maximum blob size at the chain ends. Up to a factor $A$ the
minimum blob size $\xi_0$ occurring in the 
middle of the chain is equal to $\xi_{\mbox{\scriptsize scaling}}$
\cite{castel00}. Later on  we will use $A$ as an 
adjustable parameter for comparison with simulation data. 
Using the rescaling relations given in reference \cite{castel00}
(eqs. (36) - (39)), for a weakly charged chain of electrostatic blobs
the conditions for applying the Debye-H\"uckel approximation, which
correspond to \eqns (\ref{eq:manning_rod}, \ref{eq:Debye_restrict})
discussed above, become 
\begin{equation}
  \label{eq:manning_flex}
  \left<f\right> \frac{\lambda_B \xi_{0}}{b^2}\,  
  <\, 1 ,
\end{equation}
and
\begin{equation}
  \label{eq:Debye_restrict_flex}
  \left<f\right> \frac{\lambda_D \xi_{0}}{b^2}\, >\, 1.
\end{equation}
Thus, the Manning parameter reads now $\left<f\right> \lambda_B \,
  \xi_{0}/b^2 \sim  \left(\left<f\right>\, u^{2} \right)
  ^{1/3}$. Note that the conditions imposed by applying
  the chain under tension model, i.e., (i) a fully flexible chain with
  $\xi_{0} \gg b$ and (ii) a aligned blob chain with $\lambda_D \gg
  \xi_{0}$, are even stronger than those given in \eqns (\ref{eq:manning_flex},
  \ref{eq:Debye_restrict_flex}).  
Specifically, assuming that
\begin{equation}
  \label{eq:cond_tot}
  b \ll \xi_{0} \ll \lambda_D \ll L_p,
\end{equation}
where $L_p$ is the persistence length (for inherently flexible chains
  due to electrostatic interaction), the solution of $f(s)$ up to first
  order in $\left<f\right> \lambda_B \, \xi_{0}/b^2 $ was found to be
  very similar to \eqn (\ref{eq:chargedist_rod})
  \cite{castel00} 
  \begin{eqnarray}
    \label{eq:chargedist_chain}
  \frac{f (s)}{\left< f \right>} 
   &=&  1  + 
   \left< f \right> \frac{\tilde\lambda_B}{b} 
  \\   & &\hspace*{-1cm}\times \!\left\{
   E_1\!\left[\!\left(\!\frac{N}{2}+s\!\right)\!\frac{b}{\tilde\lambda_D}
   \!\right] 
    +  
   E_1\!\left[\!\left(\!\frac{N}{2}-s\!\right)\!\frac{b}{\tilde\lambda_D}
   \!\right]  
   -  2 \frac{\tilde\lambda_D}{N b}
   \right\},
%
\nonumber
\end{eqnarray}
with ${\tilde\lambda_B}, {\tilde\lambda_D}$ being Bjerrum and
screening lengths expressed in terms of contour length 
\begin{equation}
  \label{eq:Bjerrum_resc}
  {\tilde\lambda_B} = \frac{3\xi_0}{b}\,\lambda_B,
\end{equation}
\begin{equation}
  \label{eq:screening_resc}
  {\tilde\lambda_D} = \frac{3\xi_0}{b}\,\lambda_D.
\end{equation}
The relation between persistence length and screening length
is a controversial problem which has been discussed for
a couple of decades since the early work of Odijk, Skolnick and Fixman
\cite{odejk77,skol77}. It
is beyond the scope of the paper to go into details of that
problem. Note that for fully flexible chains, the most recent results
\cite{netz99} confirm the prediction for a chain of electrostatic
blobs originally made by Khokhlov and Khachaturian  \cite{khokh82}  
\begin{equation}
  \label{eq:persistence_KK}
  L_p^{KK} \sim \xi_{\mbox{\scriptsize scaling}} +
  \frac{\lambda_D^2}{4\, \xi_{\mbox{\scriptsize scaling}}}.  
\end{equation}
Below we will use this relation to estimate the persistence length of
the weakly charged chains considered in the simulations (see
Table 1). 
\begin{table}[h]
  \begin{center}
    \caption{Systems studied and related scaling quantities ($N=1000$,
      $u=\lambda_B/b=0.9$).}  
    \label{tab:systems}
    \begin{ruledtabular}
    \begin{tabular}[t]{dcdcddd}
    \vspace*{1mm}
      \hspace*{-.4cm}
      \rule[-4mm]{0mm}{10mm}
      \left< f \right>\hspace*{-.5cm} & $\frac{{\textstyle
      \xi}_{\mbox{\scriptsize 
      scaling}}}{{\textstyle b}}$ &\hspace*{2mm} n_b &\hspace*{-4mm}
      $\left< f \right> \frac{{\textstyle \tilde\lambda}_{\scriptstyle B}
      }{{\textstyle b}}$\hspace*{2mm} &
      \frac{{\textstyle \lambda}_{\scriptstyle D}}{\textstyle b} & 
      \frac{{\textstyle \lambda}_{\scriptstyle D}}{{\textstyle
      \xi}_{\mbox{\scriptsize scaling}}}\hspace*{-.6cm} & 
      \frac{{\textstyle L}_{\scriptstyle p}^{\scriptstyle
      KK}}{{\textstyle \lambda}_{\scriptstyle D}}\hspace*{-.4cm} \\[1ex] \hline
      \rule{0mm}{3.mm}
      & & & & 16 & 1.8 & 1.5 \hspace*{.5cm}\\ \cline{5-7}
      \rule{0mm}{3.mm}\hspace*{.1cm}0.040 & 8.9 & 13 &
      \hspace*{-4mm}0.32 & 64 & 7.2 & 
      3.7 \\ \cline{5-7} 
      \rule{0mm}{3.mm}& & & & 256 & 28.8 & 16.7 \\ \hline
      \rule{0mm}{3.mm}& & & & 16 & 3.0 & 1.8 \\ \cline{5-7}
      \rule{0mm}{3.mm}0.083 & 5.4 & 34 & \hspace*{-4mm}0.40 & 64 &
      11.9 & 6.3 \\ \cline{5-7} 
      \rule{0mm}{3.mm}& & & & 256 & 47.4 & 25.0 \\ \hline
      \rule{0mm}{3.mm}& & & & 16 & 3.9 & 2.2 \\ \cline{5-7}
      \rule{0mm}{3.mm}0.125 & 4.1 & 58 & \hspace*{-4mm}0.46 & 64 &
      15.6 & 8.3 \\ \cline{5-7} 
      \rule{0mm}{3.mm}& & & & 256 & 62.4 & 30.9 \\
    \end{tabular}
    \end{ruledtabular}
  \end{center}
\end{table}
\vspace*{-.5cm}
\section{Simulation Model and Method}
\label{sec:simmodel}
In the simulation the polyelectrolyte is represented by a freely jointed
bead-spring chain, the realization of which we chose as close as possible to
the model used in theory. Along the chain the $N$ monomers are
connected to their neighbors by a harmonic bond potential 
\begin{equation}
  \label{U_SPRING}
  U_{\mbox{\scriptsize bond}} = \frac{3}{2}\,k_B T \sum_{n=1}^{N-1}\, \frac 
      {({\bf r}_{n+1} - {\bf r}_{n})^2}{b_0^2} ,
\end{equation}
with ${\bf r}_{n}$ being the position of bead $n$ and $b_0$ is the
(bare) average bond length, henceforth set equal to one. The thermal energy
is $k_B T$. For convenience, we use $k_{B}T=1$.
All $N_c$ (negatively) charged monomers interact via the
Debye-H\"uckel potential  
\begin{equation}
 U_{\mbox{\scriptsize DH}} = \frac{k_BT }{2}\sum_{n\neq m=1}^{N_c}\,
 \frac{\lambda_B}{r_{nm}}\exp\left(-\frac{r_{nm}}{\lambda_D}\right)\,,
 \label{DHpot}
\end{equation}
where the Debye screening length $\lambda_D$ is an input
parameter. The exponentially decaying interaction enables the
introduction of a cutoff which we chose as $  \lambda_c = 5 \,
\lambda_D $. In the case of weakly charged polyelectrolytes in a
$\theta$ solvent, we study in the paper, average bond length as well
as bond length distribution are only slightly effected by the
electrostatic repulsion between charged monomers. Thus, the average
bond length remains $b \approx b_0 = 1$.  
For water at room temperature, the Bjerrum length which gives the strength of
the Coulomb interaction is about \mbox{7.14 {\AA}}. To avoid problems with
counterion condensation, and to ensure that we work in a parameter
range where the theory discussed above can be applied, we set the length scale
by  $u = \lambda_B/b = 0.9$. Then the Manning parameter of a chain of
electrostatic blobs (see \eqn (\ref{eq:manning_flex})$\,$)  obeys
$\left<f\right> \lambda_B \xi_0 / b^2\, \sim \, \left(\left<f\right> u^2
\right)^{1/3} < 1$ by definition. With this setting of the length
scale one has $b \approx 8$ {\AA}.  Hence, the polyelectrolyte chain
is modeled on a coarse grained level where one bead corresponds to a few
chemical monomers. In the simulations reported here we consider a
chain which consists of $N=1000$ beads $N_c$ of which carry an elementary
charge. Because we study an annealed chain in a (semi-)grand canonical
ensemble, $N_c$ is not constant but only its average value $\left< N_c
\right>$ is fixed by the chemical potential $\mu$. The
chemical potential is chosen to result with an average degree 
of dissociation $ \left< f \right> = \left< N_c \right> / N$ 
ranging from 1/25 to 1/8. The Debye screening length $\lambda_D$ is
varied over a range from 16 to 256 average bond lengths. With the
setting of length scale introduced above such screening lengths
correspond to salt concentrations from $10^{-3}$mol/L to $10^{-5}$mol/L.  
Table 1 shows the systems studied by simulations together with the
scaling quantities being of interest with respect to the conditions
given in \eqn (\ref{eq:cond_tot}). One can see that the polyelectrolytes
considered in the simulations at least fulfill the inequality
relations which ensure that the chains reach the asymptotic regime
assumed in theory. The conditions $n_b \gg 1$ and $\left< f \right>
{\tilde\lambda_B}\, /b \ll 1$ discussed in Section 2 are also
reasonably fulfilled. 
 
Equilibrium properties of the polyions are studied by standard
Metropolis Monte Carlo (MC) \cite{metro53} simulation. In order to
guarantee the 
equilibration of both long and short length scales, bond angles and
bond lengths we combine two different configurational MC moves: (i) A
pivot move where a monomer is chosen at random and the
subsequent part of the chain is randomly rotated around that
monomer, and (ii) a local displacement move.  
The pivot algorithm was shown to be a highly efficient way to
sample the phase space in a single chain model with fixed bond lengths
\cite{Lal69,Madras88S}. In the case of our freely jointed bead-spring
model we found that the most efficient simulation route is a 
1:1 mixing of the two MC moves. The correlation time of the mean square
end-to-end distance has been checked to be of the order of a few tens
of MC steps. Thus, in any case we have correlation times below 0.1
Monte Carlo steps per monomer (MCM).

\begin{figure}[b]
  \rule{0mm}{3mm}
  \includegraphics[height=7.5cm,angle=-90]{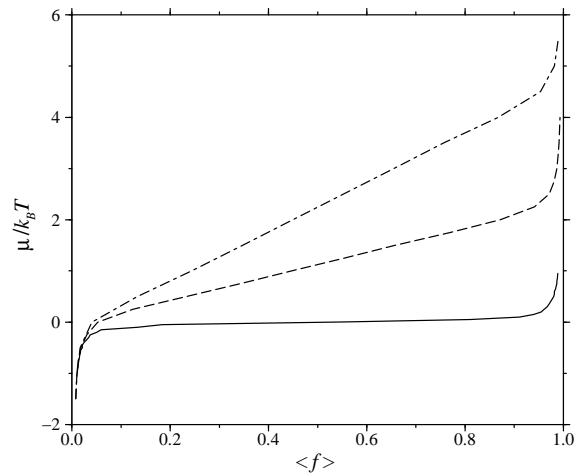}
  \caption{\label{fig:titrating} Chemical potential $\mu$ versus average
    degree of dissoziation $\left< f \right>$ for a completely
    extended rigid chain at different Bjerrum lengths:
    $\lambda_B=0$ (solid line), $\lambda_B=0.5b$ (dashed),
    $\lambda_B=1.0b$ (dot-dashed). Simulation results with
    $N=100,\, \lambda_D=10b$. 
}
  \rule{0mm}{3mm}
\end{figure}
In order to simulate annealed polyelectrolytes the MC simulation is
performed in a semi-grand canonical ensemble where the chain is in
contact with a reservoir of charges of fixed chemical potential
$\mu$. Additionally to the configurational MC moves introduced above
the algorithm is completed with a charge move by which the charge state
of a randomly chosen  monomer is switched. Again a 1:1:1 combination of the
three different MC moves was found to be the most efficient
choice. The energy change of a complete MC move reads
\begin{equation}
  \label{eq:MCenergy}
  \Delta E = \Delta E_c \pm \mu ,
\end{equation}
where $\Delta E_c$ is the change in configurational energy due to
$U_{\mbox{\scriptsize bond}}$ and $U_{\mbox{\scriptsize DH}}$. The
plus sign is used when the monomer is to be neutralized (protonated)
and the minus sign when it is to be charged (deprotonated). 
\fig \ref{fig:titrating} shows the dependence between $\mu$ and
average degree of dissociation  
$\left< f \right>$  at different strength of 
Coulomb interaction, obtained by simulating a fully extended (rigid)
polyelectrolyte.  At vanishing interaction we have the ideal curve
of isolated monomers. According to \eqn (\ref{eq:pHmu})
$\mu/k_BT$ equals $ \text{pH} - \text{p}K_a^0$. Thus, \fig
\ref{fig:titrating} represents the titration curves of the specific
model. It might seem rather academic to study the behavior at
different Bjerrum lengths. However, the important dimensionless
parameter giving the strength of Coulomb interaction is
$u=\lambda_B/b$. Hence, varying the average distance between ionizable
groups $b$, which can be easily done during polyelectrolyte synthesis,
one can succeed with a similar tuning of the strength of interaction as by
varying $\lambda_B$.

\begin{table}[b]
  \begin{center}
    \caption{Conformational properties of the systems studied: end-to-end
      distance $R$, radius of gyration $R_g$ and shape factor
      $R^2/R_g^2$ ($N=1000$, $u=\lambda_B/b=0.9$).}  
    \label{tab:properties}
    \begin{ruledtabular}
    \begin{tabular}[t]{ddccc}
    \vspace*{1mm}
      \rule[-4mm]{0mm}{9mm}
      \left< f \right>\hspace*{-.5cm} & \frac{{\textstyle
      \lambda}_{\scriptstyle D}}{\textstyle b} &  $\frac{{\textstyle
      R}}{\textstyle b}$ &  \hspace*{.3cm}$\frac{{\textstyle
      R}_{\scriptstyle g}}{\textstyle b} $& 
      $ \frac{{\textstyle
      R}^{\scriptstyle 2}}{{\textstyle
      R}_{\scriptstyle g}^2}$\hspace*{.8cm}  \\[1ex] \hline
      \rule{0mm}{3.mm}
      & 16 & 62.8$\pm$0.3 & \hspace*{.3cm}23.6$\pm$0.1 &
      7.1$\pm$0.1\hspace*{.8cm} \\ 
      \cline{2-5} 
      \rule{0mm}{3.mm}
      \hspace*{.5cm}0.040 & 64 & 75.2$\pm$0.3 & \hspace*{.3cm}27.3$\pm$0.1 &
      7.6$\pm$0.1\hspace*{.8cm} 
      \\ \cline{2-5} 
      \rule{0mm}{3.mm}
      & 256 & 77.4$\pm$0.3 & \hspace*{.3cm}28.0$\pm$0.1 &
      7.6$\pm$0.1\hspace*{.8cm} \\ \hline 
      \rule{0mm}{3.mm}
      & 16 & 98.0$\pm$0.4 & \hspace*{.3cm}35.5$\pm$0.1 &
      7.6$\pm$0.1\hspace*{.8cm} \\ 
      \cline{2-5} 
      \rule{0mm}{3.mm}
      0.083 & 64 & 132.9$\pm$0.4 & \hspace*{.3cm}45.5$\pm$0.1 &
      8.5$\pm$0.1\hspace*{.8cm} \\ \cline{2-5} 
      \rule{0mm}{3.mm}
      & 256 & 143.5$\pm$0.3 & \hspace*{.3cm}48.5$\pm$0.1 &
      8.8$\pm$0.1\hspace*{.8cm} \\ \hline 
      \rule{0mm}{3.mm}
      & 16 & 129.3$\pm$0.4 & \hspace*{.3cm}46.1$\pm$0.1 &
      7.9$\pm$0.1\hspace*{.8cm} \\ 
      \cline{2-5} 
      \rule{0mm}{3.mm}
      0.125 & 64 & 184.5$\pm$0.4 &\hspace*{.3cm} 61.5$\pm$0.1 &
      9.0$\pm$0.1\hspace*{.8cm} \\ \cline{2-5}  
      \rule{0mm}{3.mm}
      & 256 & 206.2$\pm$0.4 &\hspace*{.3cm} 67.4$\pm$0.1 &
      9.4$\pm$0.1\hspace*{.8cm} \\ 
    \end{tabular}
    \end{ruledtabular}
  \end{center}
\end{table}
\begin{figure}[top]
  \rule{0mm}{3mm}
  \includegraphics[width=7.5cm]{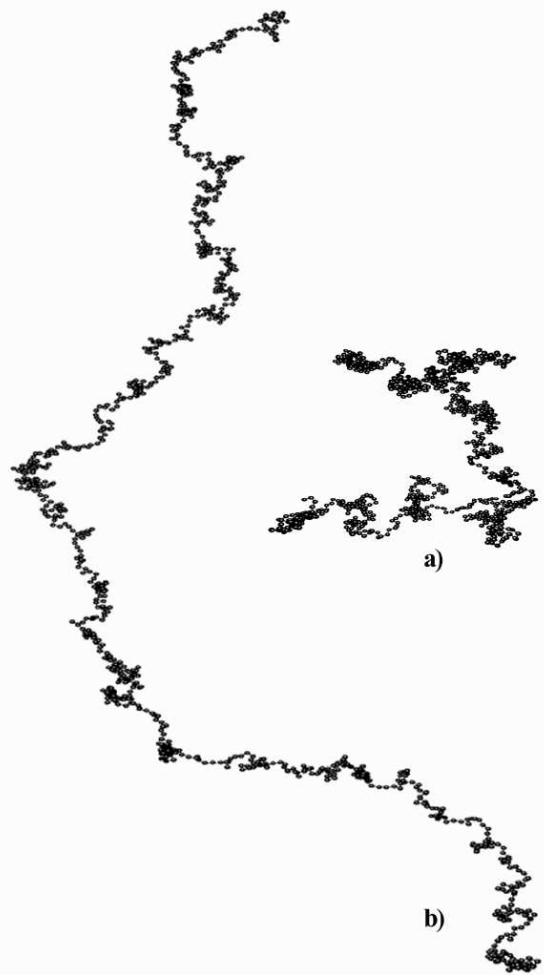}
  \caption{\label{fig:snap} Simulation snapshots of a partially charged chain
    ($N=1000$, regularly quentched distribution with $f=1/8$,
    $\theta$ solvent) at a) $\lambda_D=b$ and b) $\lambda_D=500\,b$.
} 
  \rule{0mm}{3mm}
\end{figure}
Exploring different starting configurations, specifically a random one
and the completely stretched one, the equilibration time of the chain
was estimated to be $\tau_{equ}\approx 500$ MCM. To ensure sufficient
relaxation, first the simulation is run for a time of at least 160
$\tau_{equ}$ which are about $80\cdot 10^6$ MC steps. After
reaching equilibrium, ensemble-averaged chain properties are taken as
averages over at least $1.6\cdot 10^4$ MCM, which corresponds to at
least $1.6 \cdot 10^5$ renewal times. Note that we consider one MC
step to be completed when the three partial steps (Pivot move, local move,
charge move) have been accepted. In this way the results on relevant
quantities were found to be reproducible within a few percent.

Note a significant dependence of the acceptance rate on $\left< f \right>$ and
$\lambda_D$. We estimated it to range from about 60\% for $\left< f
\right>=0.040,\, \lambda_D=16\,b$ down to about 20\% for $\left< f
\right>=0.125,\, \lambda_D=256\,b$. Due to this difference CPU times between 16
hours and 9.5 days were necessary for performing simulation runs on
Compaq Alpha machines with EV67/667MHz processors.             
\section{Simulation Results and Comparison with Theory}
\label{sec:results}
Table \ref{tab:properties} shows conformational properties of the
chains studied by simulations. As expected the stretching of the chains is
growing with increasing charge fraction $\left< f \right>$ and/or
screening length $\lambda_D$. The largest shape factor
$R^2/R_g^2$ we obtain is 9.4. Remember that it ranges from 6 for a
Gaussian chain up to 12 in the case of a rigid rod. For the largest
screening length $\lambda_D = 256 \, b$, we have always
$\lambda_D > R$, i.e., the electrostatic interaction is almost
unscreened. Note that the effect of fluctuations in the charge distribution
on large-scale mean statistical properties of free polyelectrolyte
chains in dilute solutions is weak. In \fig \ref{fig:snap} two typical
simulation snapshots of a partially charged chain are shown where
unite charges are regularly distributed with $ f=1/8 $. The qualitative
difference between the two snapshots, i.e. the polyelectrolyte effect,
is not effected by charge annealing. In the high
salt case a), the Coulomb interaction is almost completely screened
($\lambda_D = b$) and the configuration looks quite similar to a
swollen coil in a good solvent. On the other hand, in the weak
screening case b)  we have an almost linear chain of blobs. At short
length scales, i.e. inside the blobs, there are coiled regions. At
large length scales the blobs appear to build a strongly elongated
chain the transverse extension of which is considerably smaller than
the longitudinal one. An appropriate quantity that  
describes quantitatively the structure of the chain at all length
scales is the single chain structure factor or form factor. Here we
calculate the spherically averaged structure factor
\begin{equation}
  \label{eq:structure}
  S(q) = \left\langle \,\,\left\langle \frac{1}{N} \left|
        \sum_{n=1}^N \exp(i {\bf q} \cdot{\bf r}_n)
        \right|^2 \right\rangle_{|{\bf q}|}\right\rangle.
\end{equation}
\begin{figure}[b]
  \rule{0mm}{3mm}
  \includegraphics[width=7.5cm,angle=-90]{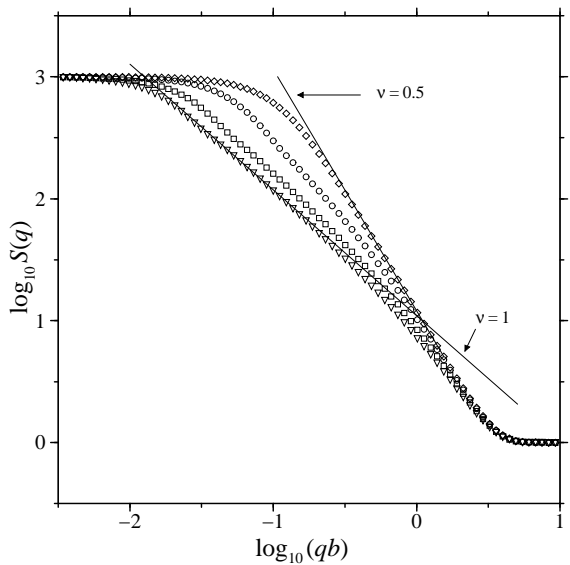}
  \caption{\label{fig:structure} Spherically averaged structure factor for
    partially charged chains ($N=1000$, annealed charge
    distribution). Simulation results at varying degree of charging
    and screening length: $\left< f  
    \right>=0.040, \lambda_D=16\,b$ (circles); $\left< f \right>=0.083,
    \lambda_D=64\,b$ (squares); $\left< f \right>=0.125,
    \lambda_D=256\,b$ (triangles). Additionally the result of an ideal
    chain is plotted (diamonds). The thin lines indicate asymptotic
    scaling laws. 
}  
  \rule{0mm}{3mm}
\end{figure}
From the theory of uncharged polymers we know that the structure
factor scales as $S(q) \sim q^{-1/\nu}$ in the range $2\pi/R < q <
2\pi/b$ with $\nu$ being the universal exponent for the mean extension
of the chain $R\sim N^\nu$. The ideal chain and good solvent chain
values of $\nu$ are 1/2 and 0.588 ($\approx$ 3/5), respectively.
\begin{figure}[b]
  \rule{0mm}{3mm}
  \includegraphics[width=7.5cm]{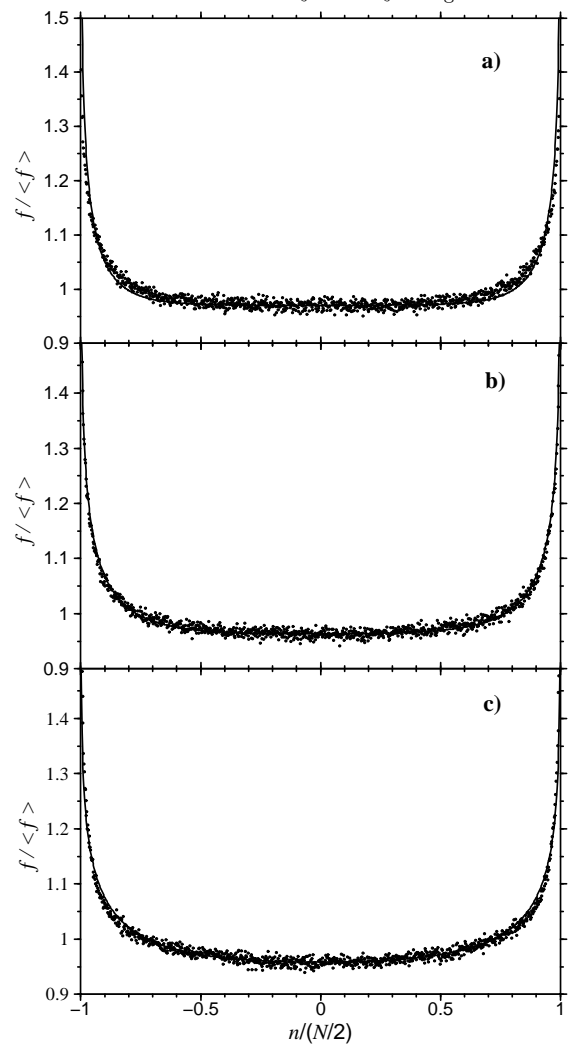}
  \caption{\label{fig:charge} Equilibrium charge distribution on annealed
    partially charged polyelectrolytes  ($N=1000$). Simulation results
    (symbols) and theoretical predictions (lines) at varying degree of
    charging and screening length:
    a) $\left< f \right>=0.040, \lambda_D=16\,b$, 
    b) $\left< f \right>=0.083, \lambda_D=64\,b$, 
    c) $\left< f \right>=0.125, \lambda_D=256\,b$.
}  
\end{figure}
\begin{table}[t]
  \begin{center}
    \caption{Fitting of charge distribution $f(s)$ with the charge
      density in the middle of the chain $f(0)$. Values of
      \mbox{parameter $A$.} }   
    \label{tab:fitting}
    \begin{ruledtabular}
    \begin{tabular*}{10cm}[t]{ddc}
    \vspace*{1mm}

%
      \hspace*{1cm}{\textstyle \lambda}_{\scriptstyle D}/{\textstyle
      b}\hspace*{-1cm} & 
      \left< f \right>\hspace*{-.5cm} & $10\times A$\hspace*{1cm}  \\[1ex] \hline
      & 0.040 & \hspace*{-.9cm}1.9 \\ 
      \hspace*{0.5cm}16\hspace*{-.7cm} & 0.083 & \hspace*{-.9cm}1.8 \\ 
      & 0.125 & \hspace*{-.9cm}1.6 \\ \hline
      & 0.040 & \hspace*{-.9cm}1.3 \\ 
      \hspace*{0.5cm}64\hspace*{-.7cm} & 0.083 & \hspace*{-.9cm}1.2 \\ 
      & 0.125 & \hspace*{-.9cm}1.2 \\ \hline
      & 0.040 & \hspace*{-.9cm}0.89 \\ 
      \hspace*{0.5cm}256\hspace*{-.7cm} & 0.083 & \hspace*{-.9cm}0.80 \\ 
      & 0.125 & \hspace*{-.9cm}0.76 \\ 
    \end{tabular*}
    \end{ruledtabular}
  \end{center}
\end{table}
In \fig \ref{fig:structure} $S(q)$  
is plotted for three different systems with annealed charge
distribution studied by simulation: (i) At
the minimum extension of the chain, (ii) at a mean one and (iii) at the 
maximum one (see Table \ref{tab:systems}). For large degree of
charging and not too short screening lengths linear scaling with $\nu
\approx 1$ is reached at large length scales. On short scales we have
an almost ideal random  
coil behavior. This is exactly the structure implied by the
theory. Thus, in contrast to the previous simulations by Berghold {\it
  et al.} \cite{berg97} now we expect a rather good agreement between 
simulation data and theory. In \fig \ref{fig:charge} we show the corresponding
charge distributions together with the theoretical predictions following from
\eqn (\ref{eq:chargedist_chain}). We remember that the theory contains a
free parameter $A$ which sets the relation between the scaling theory blob
size $\xi_{\mbox{\scriptsize scaling}}$ and blob size in the middle of a
finite chain $\xi_0 = A\,\xi_{\mbox{\scriptsize scaling}}$. We fit A
to obtain best agreement of the charge density at the middle of the
chain where it forms an extended plateau. The resulting values of $A$
are given in Table \ref{tab:fitting}. In doing so, the
agreement between simulation data and theoretical prediction is indeed almost
perfect for the case where all the three inequalities given in \eqn
(\ref{eq:cond_tot}) are fulfilled (case b{\footnotesize
  )} in \fig \ref{fig:charge}). For decreasing charging (case
a{\footnotesize )} in \fig \ref{fig:charge}), the size of
electrostatic blobs grows and becomes of the same order of
magnitude as screening length (see Table \ref{tab:systems}). Then the
interaction is too strongly screened to align the blob chain. At $\left< f
\right>=0.040,\, \lambda_D=16\,b$ the structure factor clearly shows
(see \fig \ref{fig:structure}) that the chain does not reach the
asymptotic regime with $\nu \approx 1$ assumed in the theoretical
model. Because of the stronger coiling a longer part of the chain is
packed within the range of a screening length which sets the spatial scale of
the charge inhomogeneity. Thus, in agreement with simulation data we
expect the charge to be accumulated in a larger chain section than
predicted by theory. For strong charging (case c{\footnotesize )} in
\fig \ref{fig:charge}), the expansion parameter 
$\left< f \right> {\textstyle \tilde\lambda}_{\scriptstyle B} /{\textstyle b}$
becomes too large to justify a first order perturbational treatment
used in the theoretical approach discussed in \sect \ref{sec:theory}. 
At this point we have to note that the fitting procedure of $A$
described above is actually more complex than simply finding the right
pre-factor of $\xi_{\mbox{\scriptsize scaling}}$ in the case of finite
chain length. We remember that due to the restrictions of a
first order theory the normalization of the charge distribution $f(s)$
given by \eqns  (\ref{eq:chargedist_rod}, \ref{eq:chargedist_chain})
is correct in the limit $N\gg\lambda_{\scriptstyle B}/b$ or
$N\gg{\tilde\lambda}_{\scriptstyle B}/b$, respectively. In \rrf 
\cite{berg97} it has been shown that the first order corrections 
due to finite size effects give an additive constant in $f(s)$. Thus,
the fitting of the free parameter $A$ corrects not only for the
unknown pre-factor of $\xi_{\mbox{\scriptsize scaling}}$ but also
for the higher order terms neglected in theory. However, doing so
we obtain an almost perfect agreement between simulation data and theoretical
predictions for the charge accumulation at chain ends within a length
of the order of the Debye-H\"uckel screening length.
From \eqns (\ref{eq:chargedist_chain}), (\ref{eq:Bjerrum_resc}) we know
that, in first order theory, the amplitude of charge
inhomogeneity is proportional to the rescaled strength of 
interaction, i.e., among others it is proportional to $\xi_0$. On the
other hand, Table \ref{tab:fitting} shows that the relation between
$\xi_0$ and $\xi_{\mbox{\scriptsize scaling}}$ is monotonously
reduced with growing $\lambda_{\scriptstyle D}$ and/or $\left< f
\right> $.  Thus, the larger screening length and degree of charging,
the stronger becomes the correction of the (straightforward)
first order result. Actually this tendency is in complete agreement with the
behavior we obtained for rigid rods without introducing any fitting
parameter \cite{berg97}. With growing  $\lambda_{\scriptstyle D}$ and
$f$ the first order theory overestimates the inhomogeneity, but this error
is compensated (at least partially) by the first correction term.

\section{Conclusion}
\label{sec:conclu}
In this paper we reported (semi-)grand canonical Monte Carlo
simulations of annealed weakly charged polyelectrolytes.  In a fairly
wide parameter range, i.e., not too large mean charge densities $\left<
  f \right> < b/{\tilde\lambda}_{\scriptstyle B}$ and screening lengths
in the range $\xi_{\mbox{\scriptsize scaling}} <
{\tilde\lambda}_{\scriptstyle D}< Nb$ 
we find a quite good quantitative agreement between 
simulation data and the results of the linearized theory recently
proposed by Castelnovo {\it et al.} \cite{castel00}. It would be
interesting to have experimental evidence for the accumulation of
charge at chain ends we obtain within a region of the 
order of the screening length. In order to being able to compare
simulation data with theory we have restricted the simulations to
weakly charged polyelectrolytes ($\left< f \right> \le
0.125$). However, from previous simulations \cite{berg97} it is known
that the maximum of the overcharging at chain ends appears close to
$\left< f \right> \approx 0.5$ where the charge density at the
ends can become about 50\% higher than its mean value $\left< f
\right>$. In this case it should be possible to see experimentally the
effect of the charge inhomogeneity on processes dominated by
end-effects such as, e.g., self-assembly of weakly charged micelles
\cite{vdS97} and adsorption on charged surfaces
\cite{fleer93}. But, such a charge density is clearly outside the
range where the first order theory is valid. Nonetheless the
restricted theory, now for the first time proved by simulation data,
is a step towards a more precise understanding of the more complex systems
mentioned above. 
\begin{acknowledgments}
 We thank M.~Castelnovo for helpful discussions and R.~Netz for
 critical reading of the manuscript.
\end{acknowledgments}

\end{document}